\documentclass[aps,prl,twocolumn,superscriptaddress,nofootinbib]{revtex4-2}

\usepackage{amsmath,amssymb,graphicx,bm}
\usepackage{braket}
\usepackage{comment}
\usepackage{xcolor}
\usepackage{tikz}
\usepackage{hyperref}
\usepackage{quantikz}
\DeclareMathOperator{\diag}{diag}
\DeclareMathOperator{\tr}{tr}
 
\newcommand{\RE}{\mathrm{Re}}
\newcommand{\IM}{\mathrm{Im}}

\begin{document}
	
	\title{Symmetry tests for cyclic groups with quantum linear optics}
	
	\author{Carlos Navas-Merlo}
    \email{carlos.navas@uva.es}
	\affiliation{Departamento de Teoría de la Señal y Comunicaciones e Ingeniería Telemática, Universidad de Valladolid, Paseo Bel\'en 15, 47011, Valladolid, Spain.}

	\author{Juan Carlos García Escartín}
    \email{juagar@uva.es}
	\affiliation{Departamento de Teoría de la Señal y Comunicaciones e Ingeniería Telemática, Universidad de Valladolid, Paseo Bel\'en 15, 47011, Valladolid, Spain.}
	\affiliation{Laboratory for Disruptive Interdisciplinary Science (LaDIS), Universidad de Valladolid, 47011, Valladolid, Spain.}
	\date{\today}

	\begin{abstract}
	       Testing quantum states for symmetries has multiple applications in quantum state comparison or as a primitive in more complex quantum algorithms. We present a method that determines whether an input photonic state is invariant under the action of a cyclic group defined by an operator $\hat{S}$ with eigenvalues which are roots of unity. The results generalize previously known circle and SWAP tests related to suppression laws in Fourier interferometers and offer a new test for permutations defined by a binary shift and a way to search for eigenstates in multiphoton states. Finally, we discuss the scenarios where this measurement can be used to perform Hadamard test, a fundamental primitive in variational and quantum machine-learning algorithms.
	\end{abstract}
	
	\maketitle
	
	\emph{Introduction-} Quantum linear optical systems can achieve universal quantum computation but at the cost of complex measurement assisted schemes \cite{KLM01,BBB23}. Even if, at the moment, linear optics is not as powerful as existing general-purpose noisy intermediate-scale quantum computers \cite{Pre18}, for certain tasks, like boson sampling \cite{AA11}, simple experimental systems can offer quantum advantage with respect to any classical alternative.
	
	The ability of quantum optical multiports, or quantum interferometers, to access large Hilbert spaces with standard technology makes them an interesting alternative for special-purpose quantum applications. The evolution of quantum light in linear optical system shows strong non-classical effects as exemplified by Hong-Ou-Mandel interference in a balanced beamsplitter, where two identical photons coming from different input ports always come out together suppressing coincidences \cite{HOM87}. There have been multiple generalizations to this Hong-Ou-Mandel interference \cite{LB05,TTdM10,ABG22,DKM26}. A particularly rich field is the study of the so-called ``suppression laws'' in quantum linear optics \cite{TTdM10,TTM12,Tic15,CRO16,VFI18, BS23}. In this paper we generalize previous results, particularly the work of Novo, Robbio, Galv{\~a}o and Cerf \cite{NRG26}, and show that quantum linear optics can be used to perform a wide range of symmetry tests. 
	
	Using linear optical multiports and photon counting, we can test different input states to see their projection to different invariant subspaces defined by operators whose eigenvalues are $K$th-roots of unity. These tests play a crucial role in quantum information, with some notable examples such as the SWAP test \cite{BCW01}, the circle and permutation tests \cite{BBD97,KNY08} and more general symmetry tests \cite{LaRW23,RLaW25}. The applications include previous results for a generalized SWAP tests for an arbitrary number of photons and an optical circle test \cite{NRG26}, a XOR-permutation test, a method to find eigenvectors in large Hilbert spaces and a limited implementation of the Hadamard test, a crucial primitive in quantum algorithms and quantum machine learning \cite{AJL06,LR18,SBS20,XZL22, HMS26}. We will discuss the implementation of these primitives and comment their potential applications at the end of the paper. 
		
\emph{Quantum linear optics-} We will consider Fock states $\ket{n_1 n_2 \ldots n_m}$ describing $m$ orthogonal modes with $n_i$ photons in the $i$th mode. 

The evolution of classical electromagnetic fields through a passive linear optical device with $m$ independent ports can be characterized by an $m\times m$ unitary matrix $S$, the scattering matrix \cite{Poz04}. This can be translated into its action on single photons by defining an evolution operator $\hat{S}$ that leaves the vacuum state $\ket{\mathbf{0}}$ unchanged and with an effect
\begin{equation}
\hat{a}_i^\dag\longrightarrow \hat{S} \hat{a}_i^{\dag}\hat{S}^\dag=\sum_{j=1}^{m} S_{ji} \hat{a}_j^\dag
\end{equation}
on the creation operators for mode $i$, which gives a way to describe the action on multiple photons by surrounding each creation operator by $SS^\dag=I$ terms \cite{SGL04}. 

In the Fock basis, we can define an order for the $\mathcal{M}=\binom{n+m-1}{n}$ possible states giving all the ways to put the $n$ indistinguishable photons into $m$ separate modes and describe the action of the operator $\hat{S}$ in that Hilbert space as a unitary $\mathcal{M}\times \mathcal{M}$ matrix $U=\varphi(S)$ given by a photon homomorphism $\varphi$ taking the scattering matrix $S$ into the unitary evolution $U$ \cite{AA13}. 

The scattering matrix can be diagonalized as $S=B D B^\dag$, where $D=\diag(e^{i\phi_1},e^{i\phi_2},\ldots,e^{i\phi_m})$ is a diagonal matrix with the $m$ eigenvalues of $S$ and $B$ is a change of basis matrix that takes input states with one photon in mode $k$ to an eigenstate corresponding to eigenvalue $e^{i\phi_k}$.

For any fixed number of photons 
\begin{equation}
U=\varphi(S)=\varphi(B)\varphi(D)\varphi(B)^\dag. 
\end{equation}
For superpositions of photon numbers, we can consider a Hilbert space that is the direct sum of the spaces with $n$ photons, $\oplus_{n} \mathcal{H}_{m,n}$, and write $\hat{S}=\hat{B} \hat{D}\hat{B}^\dag$. The operator $\hat{B}$ gives the effect of the multiport $B$ on the input state, which will be different for different number of photons. We can construct a block diagonal matrix with the unitary evolution for each photon number. The diagonal matrix $D$ can be interpreted as a series of phase shifters with a phase shift $\phi_k$ for mode $k$. For more than one photon, the effect can always be written as a diagonal operator in the Fock basis so that
\begin{equation}
\hat{D}\ket{n_1 n_2 \ldots n_m}=\left(\prod_{k=1}^{m} e^{i n_k \phi_k} \right)\ket{n_1 n_2 \ldots n_m}.
\end{equation}

\emph{Discrete observables and roots of unity-} Many interesting quantities are related to the powers of operators. We consider operators $\hat{S}$ where the diagonal matrix with the eigenvalues $R=\diag(e^{i\frac{2\pi}{K}a_1},e^{i\frac{2\pi}{K}a_2},\ldots,e^{i\frac{2\pi}{K}a_m})$ only contains $K$th-roots of unity $e^{i\frac{2\pi}{K}a_i}$ for integers $0 \leq a_i \leq K-1$. We choose a fixed integer $K$. If there are different roots they can be combined choosing their least common multiple as $K$. Similarly, we take the smallest possible $K$ (if all the $a_i$ have a common factor with $K$ and the description can be simplified). 

In this setting, $R^K=I$ and we have $S^k=BR^k B^\dag$ and $S^K=I$. From there, we can define the evolution operators $\hat{S}^k$ and find out the unitary evolution matrices $U^k=\varphi(S^k)$ for different photon numbers. These matrices form a cyclic group of order $K$ that is closed under matrix multiplication, with an identity element $U^K$ and each $U^k$ having an inverse $U^{K-k}$. 

We are interested in the expectation values
\begin{equation}
X_k=\tr(\hat{S}^k\rho)
\end{equation}
for input states with density matrix $\rho$ and $k=0,\ldots,K-1$. 

For an input $\rho$, the output state for evolution through a linear optical system with scattering matrix $B^\dag$ is $\Omega=\hat{B}^\dag\rho\hat{B}$. From the cyclic property of the trace operator
\begin{equation}
X_k=\tr(\hat{S}^k\rho)=\tr(\hat{B} \hat{R}^k \hat{B}^\dag\rho)=\tr( \hat{R}^k \hat{B}^\dag\rho\hat{B})=\tr(\hat{R}^k \Omega),
\end{equation}
The operator $\hat{R}^k$ is diagonal in the Fock basis, with
\begin{align}
\hat{R}^k\ket{n_1 n_2 \ldots n_m}&=\left(\prod_{l=1}^{m} e^{i \frac{2\pi}{K}a_l n_l } \right)^k\ket{n_1 n_2 \ldots n_m}&\nonumber\\
&=e^{i\frac{2\pi}{K}k\sum_{l=1}^{m}a_l n_l}\ket{n_1 n_2 \ldots n_m}.
\end{align}

The expected values $X_k$ can therefore be computed from photon number measurements at each output mode. A general Fock state $\ket{n_1n_2\ldots n_m}$ can be described by a vector $\vec{n}=(n_1,n_2,\ldots,n_m)$ giving the photon occupation for the $m$ modes.

Let's define the function
\begin{equation}
\label{function}
f(\vec{n})=\sum_{l=1}^{m} a_l n_l \mod K
\end{equation}
from the integers $a_i$ defining the eigenvalues of $S$. For any Fock state given by $\vec{n}$, $f(\vec{n})$ returns a value from $0$ to $K-1$. 

If an input state $\rho$ goes through a linear optical system described by the scattering matrix $B^\dag$ and we measure the photon number at the $m$ outputs for the output state of density matrix $\Omega$, the probability of finding a value $j$ when computing $f(\vec{n})$ is
\begin{equation}
\label{PjProb}
P_j=\sum_{f(\vec{n})=j} p(\vec{n}),
\end{equation}
where $p(\vec{n})$ is the probability of finding the Fock state described by $\vec{n}$ when measuring $\Omega$.

Following \cite{NRG26}, we can show these probabilities $P_j$ and the expected values $X_k$ have the same information and are related by a discrete Fourier transform. This relationship between probabilities and moments can be interpreted as a discrete analogue of the characteristic function and has parallels with moment-generating functions.

Consider the discrete Fourier transform
\begin{equation}
\label{Pj}
P_j=\frac{1}{K}\sum_{k=0}^{K-1}  e^{-i\frac{2\pi}{K}jk}X_k.
\end{equation}
For the operator $\hat{R}^k$, which is diagonal in the Fock basis, we have

\begin{align}
\label{e1}
P_j&=\frac{1}{K}\sum_{k=0}^{K-1}  e^{-i\frac{2\pi}{K}jk}X_k=\frac{1}{K}\sum_{k=0}^{K-1}  e^{-i\frac{2\pi}{K}jk}\tr(\hat{R}^k\Omega)=& \nonumber\\
&=\frac{1}{K}\sum_{k=0}^{K-1}  \tr\left(e^{-i\frac{2\pi}{K}jk} e^{i\frac{2\pi}{K}k\sum_{l=1}^{m} a_l \hat{n}_l}\Omega\right)&\nonumber\\
&=\frac{1}{K}\sum_{k=0}^{K-1}  \tr\left(e^{i\frac{2\pi}{K}k(\sum_{l=1}^{m} a_l \hat{n}_l-j)}\Omega\right)& \nonumber\\
&=\frac{1}{K}\sum_{k=0}^{K-1}  \sum_{\vec{n}} p(\vec{n})e^{i\frac{2\pi}{K}k(\sum_{l=1}^{m} a_l n_l-j)}.&
\end{align}
For each state in the Fock basis $\sum_{l=1}^{m} a_l n_l-j$ is an integer constant. When that constant is an integer multiple of $K$ the exponentials become 1 and the sum is $K$. For any other values of the constant, the geometric sum over $k$ cancels out. Then,
\begin{align}
\label{e2}
P_j&=\sum_{\vec{n}} p(\vec{n})\frac{1}{K}\sum_{k=0}^{K-1}  e^{i\frac{2\pi}{K}k(\sum_{l=1}^{m} a_l n_l-j)}&\nonumber\\
&=\sum_{\vec{n}} p(\vec{n})\delta(f(\vec{n})-j),
\end{align}
where we recover the definition of $P_j$ in Eq. (\ref{PjProb}) proving the probabilities are the discrete Fourier transform of the $X_k$. The inverse Fourier transform of the $P_j$, which can be estimated from measurement, gives access to all the $X_k$.

\emph{Unaffected degrees of freedom and generalized counts-} Up to this point, we have considered linear optical systems where $S$ can act on any mode of the input separately. Sometimes, it is useful to speak of internal degrees of freedom which are not affected by the optical setup. This is the usual approach in many applications where suppression laws are used to determine how indistinguishable the photons in one system really are \cite{Tic15,Cre15}. 

These additional degrees of freedom can be variables outside the control of the experimenter, but they can also be additional ways to encode information, like using different wavelengths or non-overlapping time bins.  

The effect of these distinguishable photons going through a linear optical system with scattering matrix $S$ can be described considering each of the $m$ modes has $d$ possible paths, each corresponding to one of these degrees of freedom. For $d$ degrees of freedom that are not affected by the linear optical system, we can give a description with $M=md$ modes where the $d$ modes corresponding to each of the $m$ input modes are grouped and placed in the modes going from $(k-1)d+1$ to $kd$ for $k=1,\ldots, m$. Mode $(k-1)d+l$ carries the photons corresponding to the old $k$th port, for $1\leq k\leq m$, with an internal degree of freedom $1 \leq l \leq d$. 

\begin{figure}

\includegraphics[width=0.45\textwidth]{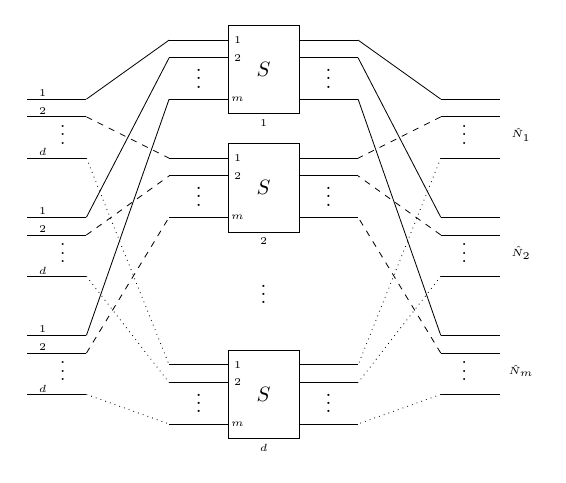}
\caption{\label{multimode}Optical setup for multimode states.}
\end{figure}

The evolution for distinguishable photons can then be reproduced with $d$ copies of the original optical system with scattering matrix $S$. We just need to direct, in order, the first modes in each group to the first copy, the second modes in each group to the second copy and so on, as shown in Figure \ref{multimode}. 

We can similarly define photon number operators 
\begin{equation}
\hat{N}_i=\sum_{k=1}^d \hat{n}_{(i,k)}
\end{equation}
for  $i=1, \ldots, m$, where $\hat{n}_{(i,k)}$ is the photon number operator for mode $(i-1)d+k$, which gives generalized counts with the total photon number in each of the $m$ groups of the $M=md$ modes.

\emph{Projections-} We focus on the probability that, after measurement, we find $f(\vec{n})=j \mod K$. These probabilities correspond to observables related to the projections of the input state $\rho$ onto different subspaces defined by the choice of the linear optical system. After postprocessing, the photon count results can be used for quantum property testing \cite{MdW16} for a wide variety of properties.

From (\ref{Pj}), $P_j$ can be written as 
\begin{align}
\label{Projectors}
P_j&=\frac{1}{K}\sum_{k=0}^{K-1}\omega^{jk}\tr(\hat{S}^k \rho)&\nonumber\\
&=\tr\left( \frac{1}{K}\sum_{k=0}^{K-1} \omega^{jk}\hat{S}^k \rho \right)=\tr(\hat{\Pi}_S^j \rho),&
\end{align}
where $\hat{\Pi}_S^j$ is a projector onto a subspace defined by $j$, the unitary matrix $S$ and $\omega=e^{-i\frac{2\pi}{K}}$. 

We can see this is actually a projector by checking $\hat{\Pi}_S^j \hat{\Pi}_S^j=\hat{\Pi}_S^j$. For operators whose eigenvalues are $K$th roots of unity, $\omega^K=1$ and $\hat{S}^K=\hat{I}$ so that
\begin{align}
\hat{\Pi}_S^j\hat{\Pi}_S^j&=\left(\frac{1}{K} \sum_{k=0}^{K-1}\omega^{jk}\hat{S}^k \right)\left(\frac{1}{K} \sum_{l=0}^{K-1}\omega^{jl}\hat{S}^l\right)&\nonumber\\
&=\frac{1}{K^2}\sum_{k=0}^{K-1}\sum_{l=0}^{K-1} \omega^{j(k+l)} \hat{S}^{k+l}&\nonumber\\
&=\frac{K}{K^2}\sum_{k=0}^{K-1}\omega^{jk}\hat{S}^{k}=\hat{\Pi}_S^j,&
\end{align}
where we use that the terms $k+l$ go exactly once through each mod $K$ value for each fixed $k$ as $l$ goes from $0$ to $K-1$. As we run through all $k$s, each possible power appears exactly $K$ times.

For input states invariant under $\hat{\Pi}_S^j$, $P_j=1$. Any measurement giving $f(\vec{n})\neq j$ proves the input state $\rho$ is not invariant under $\hat{\Pi}_S^j$. We can give multiple yes/no tests that way.

\emph{Examples-} We can now give a list of optical setups that give cyclic symmetry tests in different scenarios. In all the cases, we check the value $f(\vec{n})$ for a given $j$ and say the test has been passed if $f(\vec{n})=j$ and that the input state failed the test otherwise. These tests are asymmetric. With just one measurement result we can discard invariance under certain conditions, but we need to perform multiple measurements to increase our confidence that the state indeed satisfies the required condition. This gives direct implementations of standard quantum computing tests and generalizations. For all of the presented tests, we have run Python simulations that validate the results using the package QOptcraft \cite{GGMG23}. These simulations are available in a public repository\footnote{Code and notebooks reproducing all the examples in this paper are available at \href{https://gitlab.tel.uva.es/juagar/qoptcraft/-/tree/main/examples/Symmetry\_tests}{https://gitlab.tel.uva.es/juagar/qoptcraft/-/tree/main/examples/Symmetry\_tests}}.

\emph{Fourier multiports: state comparison} The Fourier, or symmetric, multiport gives a generalization of the balanced beamsplitter where the scattering matrix corresponds to the unitary matrix of the Quantum Fourier Transform $F$ \cite{TSJ95,TJS96,ZZH97,VD05}. Symmetric multiports have been proposed as a natural way to compare quantum states for a long time \cite{JAC04}. More recently, there has been an increased interest as this state comparison gives a way to measure distinguishability \cite{Tic15} which is valid for arbitrary photonic input states with many modes and photons \cite{NRG26}.

The Fourier multiport on $m$ input modes, with $K=m$ and the set $a_k=k-1$, gives access to the operator $\hat{C}=\hat{F}\hat{R}\hat{F}^\dag$ corresponding to a cyclic shift of the inputs in all the spatial modes by one position, with powers $\hat{C}^k$ being circular shifts by $k$ positions. This can be seen from the phase shift property of the DFT. After taking the creation operators into the transformed domain, a phase shift $e^{i\frac{2\pi}{m}ks}$ on the $k$th term corresponds to an $s$ shift in the position. 

In this case, $P_0$ tests projection into a subspace where all the states are invariant under any cyclic permutation. This corresponds to the circle test in quantum information \cite{KNY08}, which can be applied in different quantum information processing tasks, like quantum state comparison \cite{BGL24}.

This circle test is one of the possible generalizations of the SWAP test \cite{BCW01} which can be used to either determine in one measurement that two states with density matrices $\rho$ and $\sigma$ are different or, after multiple measurements, as a way to approximate $\tr(\rho\sigma)$. For two photons, Hong-Ou-Mandel interference gives a SWAP test \cite{SvE11,GC13}. When $m=2$, the symmetric test gives a generalized version of the Hong-Ou-Mandel interference for any optical state, even for inputs of different photon numbers \cite{NRG26}.

We can also use this cyclic shift test to show the tests can be restricted to subgroups. For Fourier multiports, $P_0=1$ for all states that are invariant under any cyclic permutation. But we can also define functions $f^q(\vec{n})$ replacing each $a_i$ by $q\cdot a_i$ (which can be computed from the same measurement results). This is equivalent to replacing the simulated multiport $S$ by $S^q$. If $q$ is coprime to $K$, the $X_k$ are still all different. We have the same operators in a different order (and each of them can generate the full matrix group). But if $\gcd(q,K)>1$ we have a smaller cyclic subgroup (in the Fourier case, we test symmetry under cyclic shifts by different steps). 

\emph{Sylvester interferometers: a test for XOR-permutation invariance} We can adapt the previous result to other families of permutations. Cyclic shifts are just one of the cyclic subgroups of the permutation group. 

Sylvester interferometers are an important family of linear optical multiports for which there are many known suppression laws \cite{Cre15,VFI18} and they can also be used for state comparison \cite{CDM18}. A Sylvester interferometer gives an alternative generalization of the balanced beamsplitter with a real scattering matrix which is only well defined for a number of modes $M=2^q$ for an integer $q$. This scattering matrix can be written recursively from
\begin{align}
H_1&=\frac{1}{\sqrt{2}}\begin{pmatrix}
1 & 1 \\
1 & -1 
\end{pmatrix} &\nonumber\\
&\ldots&\nonumber\\
H_{n}&=H_{n-1}\otimes H_{1}=H_{1}^{\otimes n},&
\end{align}
which can be computed from the tensor product $\otimes$ of the previous matrix and the first matrix $H_1$.
 
For a Sylvester multiport $\hat{H}$ with a scattering matrix $H_q$ and $2^q$ inputs, the single photon evolution
\begin{equation}
\label{HadT}
  \hat{H}\hat{a}^\dag_i \hat{H}^\dag\longrightarrow  \frac{1}{\sqrt{2^q}}\sum_{j=0}^{2^q-1} (-1)^{i\odot j} \hat{a}_j^\dag,
\end{equation}
can be interpreted as a Hadamard transform \cite{Ahm75} where $i\odot j$ is the bitwise inner product
\begin{equation}
i\odot j=\sum_{k=0}^{q-1} i_k j_k \mod 2,
\end{equation}
and the $x_k$ are the digits in the binary representation of integer $x=x_{q-1}\ldots x_1 x_0$. This can also be interpreted as the parity (or XOR) of the bitwise AND of the binary representations of $i$ and $j$.

For $B=H_{q}$ and $K=2$, we can choose a matrix $R$ with diagonal elements $(-1)^{k \odot s}$ for mode $k$ ($a_k=k\odot s$) to give a shift operator $\hat{X}_s=\hat{H}\hat{R}\hat{H}^\dag$ that first takes a creation operator $\hat{a}^\dag_i$ into its Hadamard transform, sets a phase $(-1)^{(i\oplus s)\odot j}$ for $\hat{a}^\dag_j$ with $\hat{R}$ and then takes the inverse Hadamard transform with output $\hat{a}^\dag_{i\oplus s}$, where $i \oplus s$ is the integer with a binary representation equal to the bitwise XOR of the binary representations of the integers $i$ and $s$. The total effect is a permutation taking photons from mode $i$ to a new mode defined by the permutation index $s$. Here, in contrast with the Fourier case, $K=2$ and all the permutations are their own inverse. 

We can define tests where $P_0=1$ indicates that the input state is invariant under a certain XOR-permutation, which depends on $s$. Notice that the same measurement can be used to run one test for each $s$. The linear optical system is the same and the only difference is the postprocessing.  

\emph{Eigenstate projection} Consider the eigenstates $\ket{e_l}$ of $U=\varphi(S)$, such that $U\ket{e_l}=\Lambda_l\ket{e_l}$ (for $l$ from $1$ to the $\mathcal{M}=\binom{n+m-1}{n}$ possible values). For any eigenstate with eigenvalue $\Lambda_l=\omega^{-j}$, $P_j=\tr(\hat{\Pi}_S^j\rho)=1$ and finding $j$ on measurement or not is a test for inputs in the eigenspace of eigenvalue $\omega^{-j}=e^{i\frac{2\pi}{K}j}$.

Inside the projectors in Eq. (\ref{Projectors})
\begin{equation}
\frac{1}{K}\sum_{k=0}^{K-1}\omega^{jk}U^{k}\ket{e_l}= \frac{1}{K}\sum_{k=0}^{K-1} (\omega^j\Lambda_l)^k\ket{e_l}
\end{equation}
with coefficients
\begin{equation}
\frac{1}{K}\sum_{k=0}^{K-1} (\omega^j\Lambda_l)^k =\begin{cases} 
      1 & \text{if } \Lambda_l=\omega^{-j}, \\
      0 & \text{otherwise.}
   \end{cases}
\end{equation}
The eigenvalues of $U$ are always of the form $\Lambda_{\vec{n}}=\lambda_1^{n_1}\lambda_2^{n_2}\cdots\lambda_m^{n_m}$, where $\lambda_i$ are the eigenvalues of $S$, for the different states in the Fock basis. This can be seen from the spectral decomposition of $S$. The diagonal matrix $D$ becomes $\varphi(D)$ where the phase of the eigenvalue in mode $i$ becomes multiplied by the number of photons in that mode, as it corresponds to a phase shifter imparting the same phase shift on each photon. For an alternative proof see \cite{Ark15}. For our scattering matrices with eigenvalues being $K$th roots of unity, $\omega^{-j}=e^{i\frac{2\pi}{K}j}$ covers all the possible values of the eigenvalues of $U$. 

The resulting test can be combined with variational methods to generate eigenvectors with a known eigenvalue using the estimation of $P_j$ as the evaluation function to maximize \cite{Gar24}. Notice that, while obtaining the full spectrum of $S$ might be easy, in general, it will be difficult to compute the eigenvectors of $U$. We are going from an $m\times m$ to an $\mathcal{M}\times \mathcal{M}$ matrix. Apart from the combinatorial growth with the number of modes and photons, computing the homomorphism giving the transformation from $S$ to $U$ is strongly believed to require extensive computations in the worst case, as shown by all the boson sampling results.

\subsection{Hadamard test} Imagine $S$ is its own inverse. Then $K=2$ and $S^2=I$, or equivalently $U^2=I$, with all the eigenvalues in $S=BRB^\dag$ being $\pm 1$ (a list of $a_i$ that are all either $0$ or $1$). For a pure state input $\rho=\ket{\psi}\bra{\psi}$ and a linear optical interferometer with scattering matrix $B^\dag$,
\begin{align}
P_0&=\tr(\hat{\Pi}_S^0\rho)=\tr\left(\frac{\hat{I}+\hat{S}}{2}\rho\right)=\tr\left(\frac{I+U}{2}\rho\right)&\nonumber\\
&=\frac{1+\bra{\psi}U\ket{\psi}}{2}.&
\end{align}
For $K=3$, $S^2=S^{-1}=S^\dag$ and 
\begin{align}
P_0&=\tr(\hat{\Pi}_S^0\rho)=\tr\left(\frac{\hat{I}+\hat{S}+\hat{S}^2}{3}\rho\right)=\tr\left(\frac{I+U+U^\dag}{3}\rho\right)&\nonumber\\
&=\frac{1}{3}+\frac{2}{3}\RE{\bra{\psi}U\ket{\psi}}.&
\end{align}	
 	
\begin{figure}[ht!]

	\begin{quantikz}
		\lstick{$|0\rangle$} & \qw & \qw  & \gate{H} & \ctrl{1} & \gate{H} & \qw & \qw & \meter{} \\
		\lstick{$|\psi\rangle$} & \qwbundle{n} & \qw & \qw    & \gate{U} & \qw & \qw & \qw & \qw 
	\end{quantikz}
\caption{\label{Htest}	Quantum circuit for the Hadamard test. The test is passed when the ancillary qubit on top is found to be on the $\ket{0}$ state, which happens with probability $\frac{1+\RE{\bra{\psi}U\ket{\psi}}}{2}$.}
\end{figure}
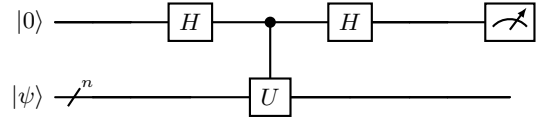

These values are related to the statistics of the quantum circuit for the Hadamard test shown in Figure \ref{Htest}. The Hadamard test can be used to estimate $\RE{\bra{\psi}U\ket{\psi}}$ and has applications in quantum algorithms and quantum machine learning, appearing in many kernels and general inner-product or gradient estimation methods \cite{AJL06,LR18,SBS20,XZL22, HMS26}. 

For $K=2$, $P_0$ has exactly the same statistics as the Hadamard test, as $U$ is Hermitian: $U^2=UU^\dag=I$ and $\bra{\psi}U\ket{\psi}=\bra{\psi}U^\dag\ket{\psi}$ is real. For $K=3$ we can estimate $\RE{\bra{\psi}U\ket{\psi}}$ after some scaling. In many applications, a complete characterization also requires an estimation of $\IM{\bra{\psi}U\ket{\psi}}$. In quantum computation, this can be done with a slightly modified circuit. For our optical description there is no direct method, but we can always combine the Hadamard test with an optical SWAP test with two equal inputs, one of them going through an interferometer with the $S$ giving $U=\varphi(S)$, to also estimate $|\bra{\psi}U\ket{\psi}|^2$. This is enough to deduce the imaginary part up to its sign.
 
As in the previous example, even if $S$ is simple, as the number of photons and modes grows, direct computation of $\RE{\bra{\psi}U\ket{\psi}}$ quickly becomes impractical, giving a potential application for non-trivial kernels in quantum machine learning with two caveats: $S$ must have $\pm 1$ or $\left\{1, e^{\pm i\frac{2\pi}{3}}\right\}$ eigenvalues and we only have access to unitaries $U=\varphi(S)$ achievable with quantum linear optics, which are limited \cite{MG17, GGM19}.
  	
We can also estimate both $\RE{\bra{\psi}U\ket{\psi}}$ and $\IM{\bra{\psi}U\ket{\psi}}$ for arbitrary matrices outside the yes/no setting of tests by computing $X_1=\tr(\hat{S}\rho)$. For pure state inputs $X_1=\bra{\psi}U\ket{\psi}$, for $U=\varphi(S)$. More generally, the characteristic function of the output probability distribution for the values of $f(\vec{n})$, $\mathcal{F}$, evaluated at 1, $\Phi(1)=E\{e^{i\mathcal{F}}\}$, gives $\tr(\hat{S}\rho)$ even for continuous eigenvalues. This requires being able to prepare multiple copies of the input state and processing the measurement results to compute the inverse Fourier transform. The proof is parallel to that in Eqs. (\ref{e1}-\ref{e2}) and the details will be presented in a future work.
 	
\emph{Summary and future lines-} We have shown that linear optical interferometers described by scattering matrices with eigenvalues that are $K$th roots of unity can be combined with photon counting to give a wide range of quantum property tests assessing the symmetries of the input states under different criteria. This includes versions of the circle and SWAP test, XOR permutation tests, eigenvector tests and a restricted Hadamard test. The described cyclic symmetry tests give new useful primitives for quantum linear optics systems and extend their potential applications in quantum machine learning or different protocols based on quantum property testing, including identity and general symmetry testing \cite{MdW16,LaRW23,BGL24}. The tests work for arbitrary input states, which can have multiple photons in separate modes, allowing for different ways to encode information. 

The description in terms of $K$th roots of unity helps to explain previous results about the measurement statistics in Fourier and Sylvester multiports and stresses previously discussed connections between spectral information and suppression laws \cite{DDW18,BS23}. Having a large number of tests can also be helpful in boson sampling validation \cite{TMB14}.

The practical realization of these tests requires the ability to build arbitrary linear optical interferometers and photon counting. There are multiple methods to realize any desired scattering matrix \cite{RZB94,CHM16,GMS18,BW21} and different experimental programmable devices that can produce any given $S$ \cite{BPC20,WSL20,ABB21,HPG22,TAdG23,WRR25}. The most technologically demanding point is precise photon counting. There exist photon-number resolving detectors for a limited resolution with different devices \cite{MG19,PLF20,CZW23,EHB23,SGB23,LSL24,ZHZ24,DZX25}. Photon counting introduces new challenges, but also gives a built-in error check against photon loss where we can discard results where the total number of photons has not been conserved. In practical applications, state preparation might also be important.
 
All the tests can be used as a one-shot property check, but also for parameter estimation. In \cite{NRG26}, the statistics for each $P_j$ in Fourier multiports were used to estimate the multitrace $\tr(\rho_1\rho_2 \ldots\rho_M)$ efficiently for $M$ different states. These statistics can also give $\tr(\hat{S}\rho)$, providing a complete Hadamard test. This will be covered in more detail in a future work. In principle, similar methods could be tried for other quantities of interest, opening an interesting line for future work.

\emph{Acknowledgements}
 J.C. Garc\'ia Escart\'in has been funded by Junta de Castilla y Le\'on (Consejer\'ia de Educaci\'on/FEDER) Project VA184P24. Financial support of the Department of Education, Junta de Castilla y Le\'on, and FEDER Funds is gratefully acknowledged (Reference: CLU-2023-1-05).
 
	\bibliographystyle{apsrev4-2}
	\bibliography{libSymmetryTestsCyclicGroups}

@inproceedings{AA11,
 author = {Aaronson, Scott and Arkhipov, Alex},
 title = {The Computational Complexity of Linear Optics},
 booktitle = {Proceedings of the 43rd Annual ACM Symposium on Theory of Computing},
 series = {STOC '11},
 year = {2011},
 isbn = {978-1-4503-0691-1},
 location = {San Jose, California, USA},
 pages = {333--342},
 numpages = {10},
 acmid = {1993682},
 publisher = {ACM},
 doi = {10.1145/1993636.1993682},
 address = {New York, NY, USA},
}

@article{AA13,
 author = {Aaronson, Scott and Arkhipov, Alex},
 title = {The Computational Complexity of Linear Optics},
 year = {2013},
 pages = {143--252},
 doi = {10.4086/toc.2013.v009a004},
 publisher = {Theory of Computing},
 journal = {Theory of Computing},
 volume = {9},
 number = {4},
 URL = {http://www.theoryofcomputing.org/articles/v009a004},
}

@article{ABB21,
  title={Quantum circuits with many photons on a programmable nanophotonic chip},
  author={Arrazola, Juan M and Bergholm, Ville and Br{\'a}dler, Kamil and Bromley, Thomas R and Collins, Matt J and Dhand, Ish and Fumagalli, Alberto and Gerrits, Thomas and Goussev, Andrey and Helt, Lukas G and others},
  journal={Nature},
  volume={591},
  number={7848},
  pages={54--60},
  year={2021},
  doi={10.1038/s41586-021-03202-1}, 
  publisher={Nature Publishing Group UK London}
}

@article{ABG22,
  title = {Extending the Hong-Ou-Mandel effect: The power of nonclassicality},
  author = {Alsing, Paul M. and Birrittella, Richard J. and Gerry, Christopher C. and Mimih, Jihane and Knight, Peter L.},
  journal = {Physical Review A},
  volume = {105},
  issue = {1},
  pages = {013712},
  numpages = {23},
  year = {2022},
  month = {Jan},
  publisher = {American Physical Society},
  doi = {10.1103/PhysRevA.105.013712}
}

@inproceedings{AJL06,
author = {Aharonov, Dorit and Jones, Vaughan and Landau, Zeph},
title = {A polynomial quantum algorithm for approximating the Jones polynomial},
year = {2006},
isbn = {1595931341},
publisher = {Association for Computing Machinery},
address = {New York, NY, USA},
doi = {10.1145/1132516.1132579},

booktitle = {Proceedings of the Thirty-Eighth Annual ACM Symposium on Theory of Computing},
pages = {427–436},
numpages = {10},
location = {Seattle, WA, USA},
series = {STOC '06}
}

@Inbook{Ahm75,
author="Ahmed, Nasir
and Rao, Kamisetty Ramamohan",
title="Walsh-Hadamard Transform",
bookTitle="Orthogonal Transforms for Digital Signal Processing",
year="1975",
publisher="Springer Berlin Heidelberg",
address="Berlin, Heidelberg",
pages="99--152",
isbn="978-3-642-45450-9",
doi="10.1007/978-3-642-45450-9_6"
}

@article{Ark15,
  title = {BosonSampling is robust against small errors in the network matrix},
  author = {Arkhipov, Alex},
  journal = {Physical Review A},
  volume = {92},
  issue = {6},
  pages = {062326},
  numpages = {5},
  year = {2015},
  month = {Dec},
  publisher = {American Physical Society},
  doi = {10.1103/PhysRevA.92.062326},
  url = {https://link.aps.org/doi/10.1103/PhysRevA.92.062326}
}

@article{BBB23,
  title={Fusion-based quantum computation},
  author={Bartolucci, Sara and Birchall, Patrick and Bombin, Hector and Cable, Hugo and Dawson, Chris and Gimeno-Segovia, Mercedes and Johnston, Eric and Kieling, Konrad and Nickerson, Naomi and Pant, Mihir and others},
  journal={Nature Communications},
  volume={14},
  number={1},
  pages={912},
  year={2023},
  doi={10.1038/s41467-023-36493-1},
  publisher={Nature Publishing Group UK London}
}

@article{BBD97,
author = {Barenco, Adriano and Berthiaume, Andr\'{e} and Deutsch, David and Ekert, Artur and Jozsa, Richard and Macchiavello, Chiara},
title = {Stabilization of Quantum Computations by Symmetrization},
journal = {SIAM Journal on Computing},
volume = {26},
number = {5},
pages = {1541-1557},
year = {1997},
doi = {10.1137/S0097539796302452}
}

@article{BCW01,
  title = {Quantum Fingerprinting},
  author = {Buhrman, Harry and Cleve, Richard and Watrous, John and de Wolf, Ronald},
  journal = {Physical Review Letters},
  volume = {87},
  issue = {16},
  pages = {167902},
  numpages = {4},
  year = {2001},
  month = {Sep},
  doi = {10.1103/PhysRevLett.87.167902}
}

@article{BGL24,
  title={Permutation tests for quantum state identity},
  author={Buhrman, Harry and Grinko, Dmitry and Lunel, Philip Verduyn and Weggemans, Jordi},
  journal={arXiv preprint arXiv:2405.09626},
  doi={10.48550/arXiv.2405.09626},
  year={2024}
}

@article{BPC20,
  title={Programmable photonic circuits},
  author={Bogaerts, Wim and P{\'e}rez, Daniel and Capmany, Jos{\'e} and Miller, David AB and Poon, Joyce and Englund, Dirk and Morichetti, Francesco and Melloni, Andrea},
  journal={Nature},
  volume={586},
  number={7828},
  pages={207--216},
  year={2020},
  doi={10.1038/s41586-020-2764-0},
  publisher={Nature Publishing Group UK London}
}

@article{BS23,
doi = {10.1088/1367-2630/acfa1e},
year = {2023},
month = {sep},
publisher = {IOP Publishing},
volume = {25},
number = {9},
pages = {093047},
author = {Bezerra, M E O and Shchesnovich, V S},
title = {Families of bosonic suppression laws beyond the permutation symmetry principle},
journal = {New Journal of Physics}
}

@article{BW21,
    author = {Bell, B. A. and Walmsley, I. A.},
    title = "{Further compactifying linear optical unitaries}",
    journal = {APL Photonics},
    volume = {6},
    number = {7},
    pages = {070804},
    year = {2021},
    month = {07},
    issn = {2378-0967},
    doi = {10.1063/5.0053421},
    url = {https://doi.org/10.1063/5.0053421},
}

@article{CDM18,
  title = {Optimal quantum-programmable projective measurement with linear optics},
  author = {Chabaud, Ulysse and Diamanti, Eleni and Markham, Damian and Kashefi, Elham and Joux, Antoine},
  journal = {Physical Review A},
  volume = {98},
  issue = {6},
  pages = {062318},
  numpages = {11},
  year = {2018},
  month = {Dec},
  publisher = {American Physical Society},
  doi = {10.1103/PhysRevA.98.062318}
  }

@article{CHM16,
author = {William R. Clements and Peter C. Humphreys and Benjamin J. Metcalf and W. Steven Kolthammer and Ian A. Walmsley},
journal = {Optica},
number = {12},
pages = {1460--1465},
publisher = {OSA},
title = {Optimal design for universal multiport interferometers},
volume = {3},
month = {Dec},
year = {2016},
url = {http://www.osapublishing.org/optica/abstract.cfm?URI=optica-3-12-1460},
doi = {10.1364/OPTICA.3.001460},
}

@article{CRO16,
  title={Suppression law of quantum states in a 3D photonic fast Fourier transform chip},
  author={Crespi, Andrea and Osellame, Roberto and Ramponi, Roberta and Bentivegna, Marco and Flamini, Fulvio and Spagnolo, Nicol{\`o} and Viggianiello, Niko and Innocenti, Luca and Mataloni, Paolo and Sciarrino, Fabio},
  journal={Nature communications},
  volume={7},
  number={1},
  pages={10469},
  year={2016},
  doi={10.1038/ncomms10469},
  publisher={Nature Publishing Group UK London}
}

@article{CZW23,
  title={A 100-pixel photon-number-resolving detector unveiling photon statistics},
  author={Cheng, Risheng and Zhou, Yiyu and Wang, Sihao and Shen, Mohan and Taher, Towsif and Tang, Hong X},
  journal={Nature Photonics},
  volume={17},
  number={1},
  pages={112--119},
  year={2023},
  doi={10.1038/s41566-022-01119-3},
  publisher={Nature Publishing Group UK London}
}

@article{Cre15,
  title = {Suppression laws for multiparticle interference in Sylvester interferometers},
  author = {Crespi, Andrea},
  journal = {Physical Review A},
  volume = {91},
  issue = {1},
  pages = {013811},
  numpages = {8},
  year = {2015},
  month = {Jan},
  publisher = {American Physical Society},
  doi = {10.1103/PhysRevA.91.013811},
  url = {https://link.aps.org/doi/10.1103/PhysRevA.91.013811}
}

@article{DDW18,
  title = {Totally Destructive Many-Particle Interference},
  author = {Dittel, Christoph and Dufour, Gabriel and Walschaers, Mattia and Weihs, Gregor and Buchleitner, Andreas and Keil, Robert},
  journal = {Physical Review Letters},
  volume = {120},
  issue = {24},
  pages = {240404},
  numpages = {6},
  year = {2018},
  month = {Jun},
  publisher = {American Physical Society},
  doi = {10.1103/PhysRevLett.120.240404}
}

@article{DKM26,
  title = {Role of Symmetry in Generalized Hong-Ou-Mandel Interference and Quantum Metrology},
  author = {Descamps, \'Eloi and Keller, Arne and Milman, P\'erola},
  journal = {Physical Review Letters},
  volume = {136},
  issue = {6},
  pages = {060807},
  numpages = {7},
  year = {2026},
  month = {Feb},
  publisher = {American Physical Society},
  doi = {10.1103/jy6g-jp7n}
}

@article{DZX25,
author = {Ding, Chaomeng and Zhang, Xingyu and Xiong, Jiamin and Xiao, You and Zhang, Tianzhu and Huang, Jia and Xu, Hongxin and Liu, Xiaoyu and You, Lixing and Wang, Zhen and Li, Hao},
title = {Photon-Number-Resolving Single-Photon Detector with a System Detection Efficiency of 98\% and Photon-Number Resolution of 32},
journal = {ACS Photonics},
volume = {12},
number = {9},
pages = {4924-4931},
year = {2025},
doi = {10.1021/acsphotonics.5c00508}

}

@article{EHB23,
  title={Resolution of 100 photons and quantum generation of unbiased random numbers},
  author={Eaton, Miller and Hossameldin, Amr and Birrittella, Richard J and Alsing, Paul M and Gerry, Christopher C and Dong, Hai and Cuevas, Chris and Pfister, Olivier},
  journal={Nature Photonics},
  volume={17},
  number={1},
  pages={106--111},
  year={2023},
  doi={10.1038/s41566-022-01105-9},
  publisher={Nature Publishing Group UK London}
}

@article{GC13,
  title = {{SWAP} test and {H}ong-{O}u-{M}andel effect are equivalent},
  author = {Garcia-Escartin, Juan Carlos and Chamorro-Posada, Pedro},
  journal = {Physical Review A},
  volume = {87},
  issue = {5},
  pages = {052330},
  numpages = {10},
  year = {2013},
  month = {May},
  doi = {10.1103/PhysRevA.87.052330}
}

@article{GGM19,
  title = {Method to determine which quantum operations can be realized with linear optics with a constructive implementation recipe},
  author = {Garcia-Escartin, Juan Carlos and Gimeno, Vicent and Moyano-Fern\'andez, Julio Jos\'e},
  journal = {Physical Review A},
  volume = {100},
  issue = {2},
  pages = {022301},
  numpages = {9},
  year = {2019},
  month = {Aug},
  publisher = {American Physical Society},
  doi = {10.1103/PhysRevA.100.022301},
  url = {https://link.aps.org/doi/10.1103/PhysRevA.100.022301}
}

@article{GGMG23,
title = {QOptCraft: A Python package for the design and study of linear optical quantum systems},
journal = {Computer Physics Communications},
volume = {282},
pages = {108511},
year = {2023},
issn = {0010-4655},
doi = {https://doi.org/10.1016/j.cpc.2022.108511},
url = {https://www.sciencedirect.com/science/article/pii/S0010465522002302},
author = {Daniel Gómez Aguado and Vicent Gimeno and Julio José Moyano-Fernández and Juan Carlos Garcia-Escartin}
}

@article{GMS18,
  title = {Simple factorization of unitary transformations},
  author = {de Guise, Hubert and Di Matteo, Olivia and S\'anchez-Soto, Luis L.},
  journal = {Physical Review A},
  volume = {97},
  issue = {2},
  pages = {022328},
  numpages = {7},
  year = {2018},
  month = {Feb},
  publisher = {American Physical Society},
  doi = {10.1103/PhysRevA.97.022328},
  url = {https://link.aps.org/doi/10.1103/PhysRevA.97.022328}
}

@article{Gar24,
  title={Finding eigenvectors with a quantum variational algorithm},
  author={Garcia-Escartin, Juan Carlos},
  journal={Quantum Information Processing},
  volume={23},
  number={7},
  pages={254},
  year={2024},
  doi={10.1007/s11128-024-04461-3},
  publisher={Springer}
}

@inproceedings{HMS26,
title={Hadamard Test is Sufficient for Efficient Quantum Gradient Estimation with Lie Algebraic Symmetries},
author={Mohsen Heidari and Masih Mozakka and Wojciech Szpankowski},
booktitle={The Thirty-ninth Annual Conference on Neural Information Processing Systems},
year={2026},
url={https://openreview.net/forum?id=PdUdzvYzoH}
}

@Article{HOM87,
  title = {Measurement of subpicosecond time intervals between two photons by interference},
  author = {Hong, C. K. and Ou, Z. Y. and Mandel, L. },
  journal = {Physical Review Letters},
  volume = {59},
  number = {18},
  pages = {2044--2046},
  numpages = {2},
  year = {1987},
  month = {Nov},
  doi = {10.1103/PhysRevLett.59.2044},
  publisher = {American Physical Society}
}

@article{HPG22,
  title={Reconfigurable continuously-coupled 3{D} photonic circuit for Boson Sampling experiments},
  author={Francesco Hoch and Simone Piacentini and Taira Giordani and Zhen-Nan Tian and Mariagrazia Iuliano and Chiara Esposito and Anita Camillini and Gonzalo Carvacho and Francesco Ceccarelli and Nicol{\'o} Spagnolo and Andrea Crespi and Fabio Sciarrino and Roberto Osellame},
  journal={npj Quantum Information},
  year={2022},
  volume={8},
  doi={10.1038/s41534-022-00568-6},
  pages={1-7}
}

@article{JAC04,
author = {Igor Jex and Erika Andersson and Anthony Chefles},
title = {Comparing the states of many quantum systems},
journal = {Journal of Modern Optics},
volume = {51},
number = {4},
pages = {505--523},
year = {2004},
publisher = {Taylor \& Francis},
doi = {10.1080/09500340408238064}
}

@article{KLM01,
  title={{A scheme for efficient quantum computation with linear optics}},
  author={Knill, E. and Laflamme, R. and Milburn, GJ},
  journal={Nature},
  volume={409},
  pages={46--52},
  year={2001},
  doi={10.1038/35051009}
}

@article{KNY08,
doi = {10.1088/1751-8113/41/39/395309},
year = {2008},
month = {sep},
publisher = {},
volume = {41},
number = {39},
pages = {395309},
author = {Kada, Masaru and Nishimura, Harumichi and Yamakami, Tomoyuki},
title = {The efficiency of quantum identity testing of multiple states},
journal = {Journal of Physics A: Mathematical and Theoretical},
}

@article{LB05,
doi = {10.1088/1367-2630/7/1/155},
year = {2005},
month = {jul},
publisher = {},
volume = {7},
number = {1},
pages = {155},
author = {Lim, Yuan Liang and Beige, Almut},
title = {Generalized Hong–Ou–Mandel experiments with bosons and fermions},
journal = {New Journal of Physics}
}

@article{LR18,
  title = {Quantum machine learning for quantum anomaly detection},
  author = {Liu, Nana and Rebentrost, Patrick},
  journal = {Physical Review A},
  volume = {97},
  issue = {4},
  pages = {042315},
  numpages = {10},
  year = {2018},
  month = {Apr},
  publisher = {American Physical Society},
  doi = {10.1103/PhysRevA.97.042315},
  url = {https://link.aps.org/doi/10.1103/PhysRevA.97.042315}
}

@article{LSL24,
    author = {Los, J. W. Niels and Sidorova, Mariia and Lopez-Rodriguez, Bruno and Qualm, Patrick and Chang, Jin and Steinhauer, Stephan and Zwiller, Val and Zadeh, Iman Esmaeil},
    title = {High-performance photon number resolving detectors for 850–950 nm wavelength range},
    journal = {APL Photonics},
    volume = {9},
    number = {6},
    pages = {066101},
    year = {2024},
    month = {06},
    issn = {2378-0967},
    doi = {10.1063/5.0204340}
}

@article{LaRW23,
  doi = {10.22331/q-2023-09-25-1120},
  title = {Testing symmetry on quantum computers},
  author = {LaBorde, Margarite L. and Rethinasamy, Soorya and Wilde, Mark M.},
  journal = {{Quantum}},
  issn = {2521-327X},
  publisher = {{Verein zur F{\"{o}}rderung des Open Access Publizierens in den Quantenwissenschaften}},
  volume = {7},
  pages = {1120},
  month = sep,
  year = {2023}
}

@article{MG17,
author = {Julio Jos\'e Moyano-Fern\'andez and Juan Carlos Garcia-Escartin},
title = "Linear optics only allows every possible quantum operation for one photon or one port",
journal = "Optics Communications",
volume = "382",
pages = "237 -- 240",
doi = "10.1016/j.optcom.2016.07.085",
year = "2017"
}

@article{MG19,
  title = {Evaluating the performance of photon-number-resolving detectors},
  author = {J\"onsson, Mattias and Bj\"ork, Gunnar},
  journal = {Physical Review A},
  volume = {99},
  issue = {4},
  pages = {043822},
  numpages = {8},
  year = {2019},
  month = {Apr},
  publisher = {American Physical Society},
  doi = {10.1103/PhysRevA.99.043822}
}

@book{MdW16,
 author = {Montanaro, Ashley and Wolf, Ronald {de}},
 title = {A Survey of Quantum Property Testing},
 year = {2016},
 pages = {1--81},
 doi = {10.4086/toc.gs.2016.007},
 publisher = {Theory of Computing Library},
 number = {7},
 series = {Graduate Surveys},
 URL = {http://www.theoryofcomputing.org/library.html},
}

@article{NRG26,
  title={Native linear-optical protocol for efficient multivariate trace estimation},
  author={Novo, Leonardo and Robbio, Marco and Galv{\~a}o, Ernesto F and Cerf, Nicolas J},
  journal={arXiv preprint arXiv:2601.14204},
  doi={10.48550/arXiv.2601.14204},
  year={2026}
}

@article{PLF20,
author = {Jan Provazn\'{i}k and Luk\'{a}\v{s} Lachman and Radim Filip and Petr Marek},
journal = {Optics Express},
number = {10},
pages = {14839--14849},
publisher = {Optica Publishing Group},
title = {Benchmarking photon number resolving detectors},
volume = {28},
month = {May},
year = {2020},
doi = {10.1364/OE.389619},
}

@book{Poz04,
  title={Microwave Engineering},
  author={Pozar, D.M.},
  isbn={9780471448785},
  edition={Fourth},
  year={2004},
  nolink={},
  publisher={Wiley}
}

@article{Pre18,
  doi = {10.22331/q-2018-08-06-79},
  url = {https://doi.org/10.22331/q-2018-08-06-79},
  title = {Quantum {C}omputing in the {NISQ} era and beyond},
  author = {Preskill, John},
  journal = {{Quantum}},
  issn = {2521-327X},
  publisher = {{Verein zur F{\"{o}}rderung des Open Access Publizierens in den Quantenwissenschaften}},
  volume = {2},
  pages = {79},
  month = aug,
  year = {2018}
}

@article{RLaW25,
author = {Rethinasamy, Soorya and LaBorde, Margarite L. and Wilde, Mark M.},
title = {Quantum computational complexity and symmetry},
journal = {Canadian Journal of Physics},
volume = {103},
number = {2},
pages = {215-239},
year = {2025},
doi = {10.1139/cjp-2023-0260}
}

@ARTICLE{RZB94,
    author = {M. Reck  and A. Zeilinger and H.J. Bernstein and P. Bertani},
    title = {Experimental Realization of Any Discrete Unitary Operator},
    journal = {Physical Review Letters},
    year = {1994},
    volume = {73},
    number = {1},
    pages = {58--61},
    day = {4},
    month = {July},
    doi={10.1103/PhysRevLett.73.58}
 }

@article{SBS20,
	author    = {Maria Schuld and Alex Bocharov and Krysta Svore and Nathan Wiebe},
	title     = {Circuit-centric quantum classifiers},
	journal   = {Physical Review A},
	volume    = {101},
	pages     = {032308},
	year      = {2020},
	doi       = {10.1103/PhysRevA.101.032308}
}

@article{SGB23,
  title = {Fast High-Efficiency Photon-Number-Resolving Parallel Superconducting Nanowire Single-Photon Detector},
  author = {Stasi, Lorenzo and Gras, Ga\"etan and Berrazouane, Riad and Perrenoud, Matthieu and Zbinden, Hugo and Bussi\`eres, F\'elix},
  journal = {Physical Review Applied},
  volume = {19},
  issue = {6},
  pages = {064041},
  numpages = {8},
  year = {2023},
  month = {Jun},
  publisher = {American Physical Society},
  doi = {10.1103/PhysRevApplied.19.064041},
  url = {https://link.aps.org/doi/10.1103/PhysRevApplied.19.064041}
}

@article{SGL04,
author = {Johannes Skaar and Juan Carlos {Garc\'{\i}a Escart\'{\i}n} and Harald Landro},
collaboration = {},
title = {Quantum mechanical description of linear optics},
publisher = {AAPT},
year = {2004},
journal = {American Journal of Physics},
volume = {72},
number = {11},
pages = {1385-1391},
url = {http://link.aip.org/link/?AJP/72/1385/1},
doi = {10.1119/1.1775241}
}

@article{SvE11,
  title = {Detecting the Drift of Quantum Sources: Not the de {F}inetti Theorem},
  author = {Schwarz, Lucia and van Enk, S. J.},
  journal = {Physical Review Letters},
  volume = {106},
  issue = {18},
  pages = {180501},
  numpages = {4},
  year = {2011},
  month = {May},
 doi = {10.1103/PhysRevLett.106.180501}
}

@article{TAdG23,
  doi = {10.22331/q-2023-08-01-1071},
  title = {20-{M}ode {U}niversal {Q}uantum {P}hotonic {P}rocessor},
  author = {Taballione, Caterina and Anguita, Malaquias Correa and de Goede, Michiel and Venderbosch, Pim and Kassenberg, Ben and Snijders, Henk and Kannan, Narasimhan and Vleeshouwers, Ward L. and Smith, Devin and Epping, J{\"{o}}rn P. and van der Meer, Reinier and Pinkse, Pepijn W. H. and van den Vlekkert, Hans and Renema, Jelmer J.},
  journal = {{Quantum}},
  issn = {2521-327X},
  publisher = {{Verein zur F{\"{o}}rderung des Open Access Publizierens in den Quantenwissenschaften}},
  volume = {7},
  pages = {1071},
  month = aug,
  year = {2023}
}

@article{TJS96,
author = {P. T\"orm\"a and I. Jex and S. Stenholm},
title = {Beam splitter realizations of totally symmetric mode couplers},
journal = {Journal of Modern Optics},
volume = {43},
number = {2},
pages = {245--251},
year = {1996},
publisher = {Taylor \& Francis},
doi = {10.1080/09500349608232738}
}

@article{TMB14,
  title = {Stringent and Efficient Assessment of Boson-Sampling Devices},
  author = {Tichy, Malte C. and Mayer, Klaus and Buchleitner, Andreas and M\o{}lmer, Klaus},
  journal = {Physical Review Letters},
  volume = {113},
  issue = {2},
  pages = {020502},
  numpages = {5},
  year = {2014},
  month = {Jul},
  publisher = {American Physical Society},
  doi = {10.1103/PhysRevLett.113.020502}
}

@article{TSJ95,
  title = {Hamiltonian theory of symmetric optical network transforms},
  author = {T\"orm\"a, P\"aivi and Stenholm, Stig and Jex, Igor},
  journal = {Physical Review A},
  volume = {52},
  issue = {6},
  pages = {4853--4860},
  numpages = {0},
  year = {1995},
  month = {Dec},
  publisher = {American Physical Society},
  doi = {10.1103/PhysRevA.52.4853}
 
}

@article{TTM12,
doi = {10.1088/1367-2630/14/9/093015},
year = {2012},
month = {sep},
publisher = {IOP Publishing},
volume = {14},
number = {9},
pages = {093015},
author = {Tichy, Malte C and Tiersch, Markus and Mintert, Florian and Buchleitner, Andreas},
title = {Many-particle interference beyond many-boson and many-fermion statistics},
journal = {New Journal of Physics}
}

@article{TTdM10,
  title = {Zero-Transmission Law for Multiport Beam Splitters},
  author = {Tichy, Malte Christopher and Tiersch, Markus and de Melo, Fernando and Mintert, Florian and Buchleitner, Andreas},
  journal = {Physical Review Letters},
  volume = {104},
  issue = {22},
  pages = {220405},
  numpages = {4},
  year = {2010},
  month = {Jun},
  publisher = {American Physical Society},
  doi = {10.1103/PhysRevLett.104.220405}
}

@article{Tic15,
  title = {Sampling of partially distinguishable bosons and the relation to the multidimensional permanent},
  author = {Tichy, Malte C.},
  journal = {Physical Review A},
  volume = {91},
  issue = {2},
  pages = {022316},
  numpages = {13},
  year = {2015},
  month = {Feb},
  publisher = {American Physical Society},
  doi = {10.1103/PhysRevA.91.022316}
}

@article{VD05,
  title = {Fourier multiport devices},
  author = {Vourdas, A. and Dunningham, J. A.},
  journal = {Physical Review A},
  volume = {71},
  issue = {1},
  pages = {013809},
  numpages = {6},
  year = {2005},
  month = {Jan},
  publisher = {American Physical Society},
  doi = {10.1103/PhysRevA.71.013809}
}

@article{VFI18,
doi = {10.1088/1367-2630/aaad92},
year = {2018},
month = {mar},
publisher = {IOP Publishing},
volume = {20},
number = {3},
pages = {033017},
author = {Viggianiello, Niko and Flamini, Fulvio and Innocenti, Luca and Cozzolino, Daniele and Bentivegna, Marco and Spagnolo, Nicolò and Crespi, Andrea and Brod, Daniel J and Galvão, Ernesto F and Osellame, Roberto and Sciarrino, Fabio},
title = {Experimental generalized quantum suppression law in Sylvester interferometers},
journal = {New Journal of Physics}
}

@article{WRR25,
  title={Scalable photonic quantum technologies},
  author={Wang, Hui and Ralph, Timothy C and Renema, Jelmer J and Lu, Chao-Yang and Pan, Jian-Wei},
  journal={Nature Materials},
  volume={24},
  number={12},
  pages={1883--1897},
  year={2025},
  doi={10.1038/s41563-025-02306-7},
  publisher={Nature Publishing Group UK London}
}

@article{WSL20,
  title={Integrated photonic quantum technologies},
  author={Wang, Jianwei and Sciarrino, Fabio and Laing, Anthony and Thompson, Mark G},
  journal={Nature photonics},
  volume={14},
  number={5},
  pages={273--284},
  year={2020},
  doi={10.1038/s41566-019-0532-1},
  publisher={Nature Publishing Group UK London}
}

@article{XZL22,
doi = {10.1088/1572-9494/ac6358},
year = {2022},
month = {may},
publisher = {IOP Publishing},
volume = {74},
number = {5},
pages = {055106},
author = {Xu, Li and Zhang, Xiao-Yu and Liang, Jin-Min and Wang, Jing and Li, Ming and Jian, Ling and Shen, Shu-qian},
title = {Variational quantum support vector machine based on Hadamard test},
journal = {Communications in Theoretical Physics}
}

@article{ZHZ24,
author = {Tianzhu Zhang and Jia Huang and Xingyu Zhang and Chaomeng Ding and Huiqin Yu and You Xiao and Chaolin Lv and Xiaoyu Liu and Zhen Wang and Lixing You and Xiaoming Xie and Hao Li},
journal = {Photonics Research},
number = {6},
pages = {1328--1333},
publisher = {Optica Publishing Group},
title = {Superconducting single-photon detector with a speed of 5\&\#x2009;\&\#x2009;GHz and a photon number resolution of 61},
volume = {12},
month = {Jun},
year = {2024},
url = {https://opg.optica.org/prj/abstract.cfm?URI=prj-12-6-1328},
doi = {10.1364/PRJ.522714},
}

@article{ZZH97,
  title = {Realizable higher-dimensional two-particle entanglements via multiport beam splitters},
  author = {\ifmmode \dot{Z}\else \.{Z}\fi{}ukowski, Marek and Zeilinger, Anton and Horne, Michael A.},
  journal = {Physical Review A},
  volume = {55},
  issue = {4},
  pages = {2564--2579},
  numpages = {0},
  year = {1997},
  month = {Apr},
  publisher = {American Physical Society},
  doi = {10.1103/PhysRevA.55.2564}
}
	
\end{document}